\documentclass{SCGE}
\usepackage[colorlinks,linkcolor=blue,citecolor=blue,urlcolor=blue]{hyperref}

\usepackage[utf8]{inputenc}
\usepackage{rotating}
\usepackage{footnote}
\newcolumntype{C}[1]{>{\centering\let\newline\\\arraybackslash\hspace{0pt}}m{#1}}
\usepackage[math]{blindtext}
\usepackage{times}
\usepackage{newtxtext,newtxmath}
\usepackage{graphicx}	
\usepackage{amsmath}	
\usepackage{amssymb}	
\usepackage{xspace}
\usepackage{bigints}
\usepackage{siunitx}

\usepackage{amsthm}
\usepackage{mathtools}
\usepackage[normalem]{ulem}
\usepackage{float}
\usepackage{color}
\usepackage[sort&compress,numbers]{natbib}
\usepackage{gensymb}

\usepackage{CJK}
\usepackage{upgreek}
\usepackage{booktabs}

\def\be{\begin{equation}}
\def\ee{\end{equation}}

\let\citedash\relax
\makeatletter \providecommand{\citedash}{\hbox{-}\penalty\@m}
\makeatother
\begin{document}
\begin{CJK*}{UTF8}{gbsn}

\begin{picture}(0,0){\rm
\put(0,-20){\makebox[160truemm][l]{\bf {\sanhao\raisebox{2pt}{.}}
Article  {\sanhao\raisebox{1.5pt}{.}}}}}
\put(0,-34){\jiuwuhao {\textcolor[rgb]{0.5,0.5,0.5}{\sf 
}}}
\end{picture}

\def\bm{\boldsymbol}

\def\dl{\displaystyle}
\def\du{\end{document}}
\def\d{{\rm d}}
\def\e{{\rm e}}
\def\i{{\rm i}}

\Year{2021} %
\Month{XXXX} %
\Vol{xx} %
\No{x} %
\BeginPage{1} %
\AuthorMark{{\rm Wei-Yang Wang et al.} }  
\DOI{} 
\ArtNo{000000}

\title[Fast Radio Burst]{Repeating Fast Radio Bursts:\\
{\em Coherent Circular Polarization by Bunches}}

\author[1,2]{Wei-Yang Wang}{Corresponding author (email: wywang\_astroph@pku.edu.cn)}%
\author[1,2]{Jin-Chen Jiang}{}%
\author[3,4]{Jiguang Lu}{}%
\author[2]{Heng Xu}{}%
\author[2]{Jiangwei Xu}{}%
\author[2]{Kejia Lee}{}%
\author[5,6]{Jifeng Liu}{}%
\author[1,2]{Renxin Xu}{Corresponding author (email: r.x.xu@pku.edu.cn)}%

\address[{\rm1}]{School of Physics and State Key Laboratory of Nuclear Physics and Technology, Peking University, Beijing 100871, P.R.China;}
\address[{\rm2}]{Kavli Institute for Astronomy and Astrophysics, Peking University, Beijing 100871, P.R.China;}
\address[{\rm3}]{CAS Key Laboratory of FAST, National Astronomical Observatories, Chinese Academy of Sciences, Beijing 100101, P.R.China;}
\address[{\rm4}]{Guizhou Radio Astronomy Observatory, Chinese Academy of Sciences, Guiyang 550025, P.R.China;}
\address[{\rm5}]{Key Laboratory of Optical Astronomy, National Astronomical Observatories, Chinese Academy of Sciences, Beijing, P.R.China;}
\address[{\rm6}]{School of Astronomy and Space Sciences, University of Chinese Academy of Sciences, Beijing, P.R.China;}

\maketitle \vspace{-3.5mm}{\footnotesize\begin{center} Received Month date, Year; accepted Month date, Year
\end{center}}\vspace*{-5mm}

\begin{center}
\rule{16.5cm}{0.4pt}
\parbox{16.5cm}
{\begin{abstract}

Fast radio bursts (FRBs) are millisecond-duration signals that are highly dispersed at distant galaxies. However, the physical origin of FRBs is still unknown.
Coherent curvature emission by bunches, e.g., powered by starquakes, has already been proposed for repeating FRBs. It has the nature of understanding narrowband radiation exhibiting time--frequency drifting.
Recently, a highly active FRB source, i.e., FRB 20201124A, was reported to enter a newly active episode and emit at least some highly circular-polarized bursts.
In this study, we revisit the polarized FRB emission, particularly investigating the production mechanisms of a highly circular polarization (CP) by deriving the intrinsic mechanism and propagative effect.
The intrinsic mechanisms of invoking charged bunches are approached with radiative coherence. Consequently,
a highly CP could naturally be explained by the coherent summation of outcome waves, generated or scattered by bunches, with different phases and electric vectors.
Different kinds of evolutionary trajectories are found on the Poincar{\' e} sphere for the bunch-coherent polarization, and this behavior could be tested through future observations.
Cyclotron resonance can result in the absorption of R-mode photons at a low altitude region of the magnetosphere, and an FRB should then be emitted from a high-altitude region if the waves have strong linear polarization.
Circularly polarized components could be produced from Faraday conversion exhibiting a $\lambda^3$-oscillation, but the average CP fraction depends only on the income wave, indicating a possibility of a highly circular-polarized income wave.
The analysis could be welcome if extremely high (e.g., almost 100\%) CP from repeating FRBs is detected in the future.
Finally, the production of a bulk of energetic bunches in the pulsar-like magnetosphere is discussed, which is relevant to the nature of the FRB central engine.

\end{abstract}}
\end{center}\vspace*{-0.6cm}

\begin{center}
\parbox{16.5cm}
{\bf\jiuhao Fast radio burst, Radiation mechanism, Neutron stars}
\end{center}

\begin{center}
{}
\end{center}

\textwidth=178truemm \textheight=236truemm

\wuhao\vspace*{1.5mm}

\renewcommand{\baselinestretch}{1.08} \baselineskip 12.2pt\parindent=10.8pt


\section{Introduction}
\label{section:1}

Fast radio bursts (FRBs) are millisecond-duration and bright transients prevailing in the universe.
Since the first discovery of FRBs \cite{Lorimer07}, the field has witnessed a rapid increase in the frontiers of observations and theories ~\cite{Cordes19,Petroff19,Zhang20,Xiao21}.
The key issue in this field is to understand the physical origin(s) of FRBs and to be still mysterious although the known samples have been enlarged up to more than hundreds and dozens of them can be reproduced \footnote{See the Transient Name Server, \href{https://www.wis-tns.org/}{https://www.wis-tns.org/}}.

FRB sources fall into two groups: repeaters and apparently non-repeating FRBs.
Although the two groups exhibit noticeable differences, whether all FRBs can repeat is still an open question \footnote{We mainly focus on repeating FRBs in this paper.}.
These bursts have dispersion measures (DMs) in excess of the Galactic values. Accordingly, their sources were localized in their host galaxies, which establish the cosmological origin of FRBs \cite{Bassa17,Chatterjee17,Marcote17,Bannister19,Ravi19,Macquart20,Marcote20}.
An intriguing spectral structure across subpulses has been found in some repeating FRBs, which is nicknamed the ``sad trombone'' drift pattern \cite{CHIME19a,CHIME19b,Hessels19,Josephy19,Caleb20,Day20,Fonseca20,Hilmarsson21b}.
A bright FRB-like burst associated with temporal coincident X-ray bursts was discovered from the Galactic magnetar SGR J1935+2154 \cite{Bochenek20,CHIME20,Mereghetti20,Li21,Ridnaia21,Tavani21}, indicating that magnetars are most likely related to FRBs.
An extremely coherent radiation mechanism is required to explain the brightness temperature ($\sim10^{35}$K or even larger) of FRBs.
Theoretical models that invoke such coherent mechanisms can generally be characterized into two categories: pulsar-like~\cite{Kumar20,Wang20,Lu20,Yang21} and gamma-ray-burst-like ones ~\cite{Metzger19,Beloborodov20,Margalit20}.

The pulsar-like mechanism is indicative of the coherent emission of charged bunches from the magnetosphere, but FRB emissions are akin to a sudden and violent trigger for the sparking process.
Some earthquake-like behaviors were found in a sequence of repeating bursts for FRB 121102, suggesting that FRB pulses may originate from sudden starquakes of a pulsar \cite{Wang18}.
A trigger event can excite the coherent bunches of charged particles, which travel farther but are seen later, emitting at low frequencies so that one can observe a downward drift pattern \cite{Wang19}.
An upward drift pattern, for instance, the two subpulse components of SGR J1935+2154, can be produced when bunches are generated at slightly different times \citep{Wang20}.
The properties of narrowband emissions can be well understood within the framework of coherent curvature radiation (CR) \cite{Yang18,Wang21}, and bursts should be extremely linearly polarized for on-beam cases.

Careful polarization measurements are potential tools to shed light on more information on the radiation mechanism, especially taking advantage of the extremely high sensitivity
of the biggest single-dish radio telescope, China's Five-hundred-meter Aperture Spherical radio Telescope (FAST)~\cite{Lu20fast,XuR2021}.
Most bursts have linear polarization (LP) fractions up to $100\%$ and present a flat polarization angle (PA) across each pulse \cite{Michilli18,Hilmarsson21}.
However, some sources, for instance, FRB 180301 and FRB 181112, exhibit quite different properties in PAs across the burst envelope, which are variables across each burst \cite{Cho20,Luo20}.
This property is reminiscent of an S-shaped PA across the burst envelope of a pulsar, which is a well-known phenomenon in some
normal radio pulsars and is explained as a rotating magnetosphere \cite{Radhakrishnan69}.
Recently, a CHIME-discovered source \cite{CHIME21}, namely, FRB 20201124A, was reported to show an active episode from April to May 2021, and hundreds of bursts were detected
in that month with a variety of radio telescopes \cite{Day21,Farah21,Herrmann21,Hilmarsson21b,Kilpatrick21,Kumar21,Law21,Ricci21,Wharton21,Xu21a}.
Most bursts from FRB 121102 are highly linearly polarized, with the flat PA across the burst envelope, and some of them exhibit a downward drift spectro-temporal structure \cite{Hilmarsson21b}.
A burst with a significant circular polarization (CP) fraction, which is up to $47\%$ \cite{Kumar21}, has never been seen before in bursts from repeating FRB sources.
Surprisingly, more bursts with CP factions of $>50\%$ have been found there, and even one with $75\%$, refreshing the previous record \cite{Xu21b}.

In this paper, we focus on demonstrating the polarization properties within the framework of a generic pulsar-like model by invoking charged bunches in the magnetosphere.
FRB emissions could intrinsically be produced using bunches of coherent CR \cite{Kumar17,Yang18}, which is explained in Section \ref{sec2}.
Propagation effects are investigated in Section \ref{sec3}.
The results are discussed and summarized in Sections \ref{sec4} and \ref{sec5}, respectively.
The convention $Q_x=Q/10^x$ in the cgs units is used throughout the paper.

\section{Coherent CR By Bunches}\label{sec2}

\begin{figure}
 \centering
 \includegraphics[width = 7.5cm]{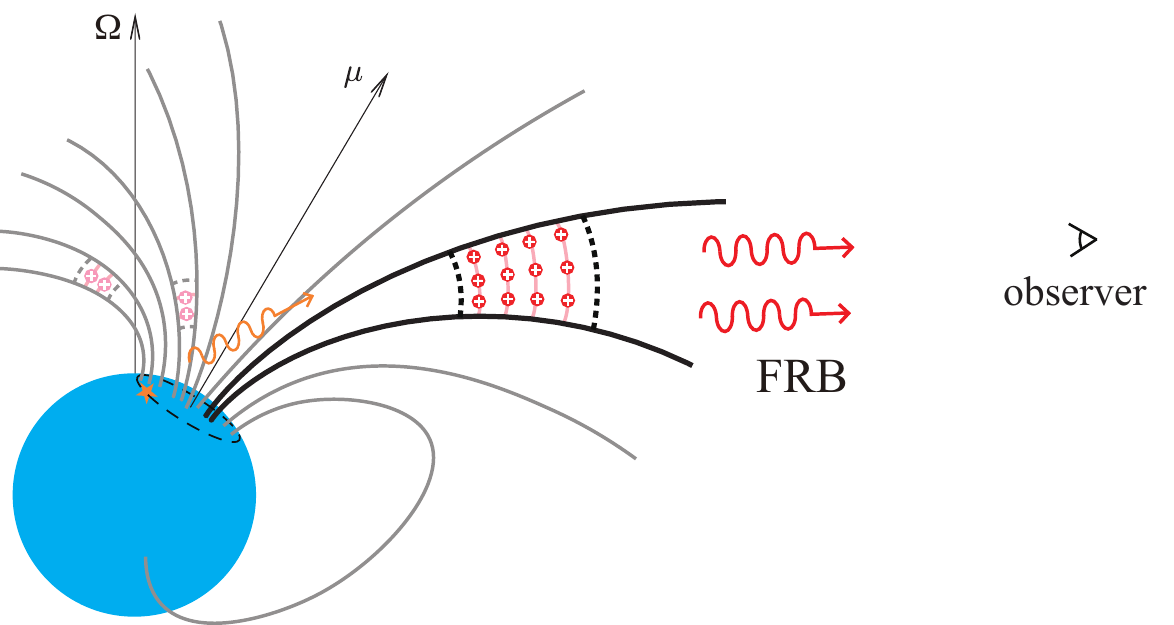}
 \caption{Generic schematic diagram of a bulk of bunches contributing to an observed FRB. The light solid red line presents the slice (i.e., a bunch), in which charged particles are emitted in the same phase. A bulk of bunches could generate an FRB emission via either a curvature (CR) radiation or inverse Compton scattering (ICS). For the ICS case, a sparking event occurs at a point (orange pentagram) above the polar cap and produces the incident waves that are scattered in bunches.}
 \label{fig:fig1}
\end{figure}

Several electron-positron plasma pairs would be created via sparking events near the inner polar cap of a neutron star \cite{RS75}.
The production of charged bunches is proposed to be formed by two-stream instabilities \cite{Usov87,Asseo98,Melikidze00} and has been simulated in open field lines \cite{Philippov20,Benacek21}.
Positive net charges are the point of interest in the following discussion.
Near the polar cap region, the trajectories of charged particles are tracked by magnetic field lines because of the rapid dump of the vertical momentum perpendicular to the field line.
These charged bunches stream outward along the magnetic field lines, emitting FRB waves via CR \cite{Kumar17,Ghisellini18,Lu18,Yang18,Wang20b,Cooper21}

\begin{figure}
 \centering
 \includegraphics[width = 7.5cm]{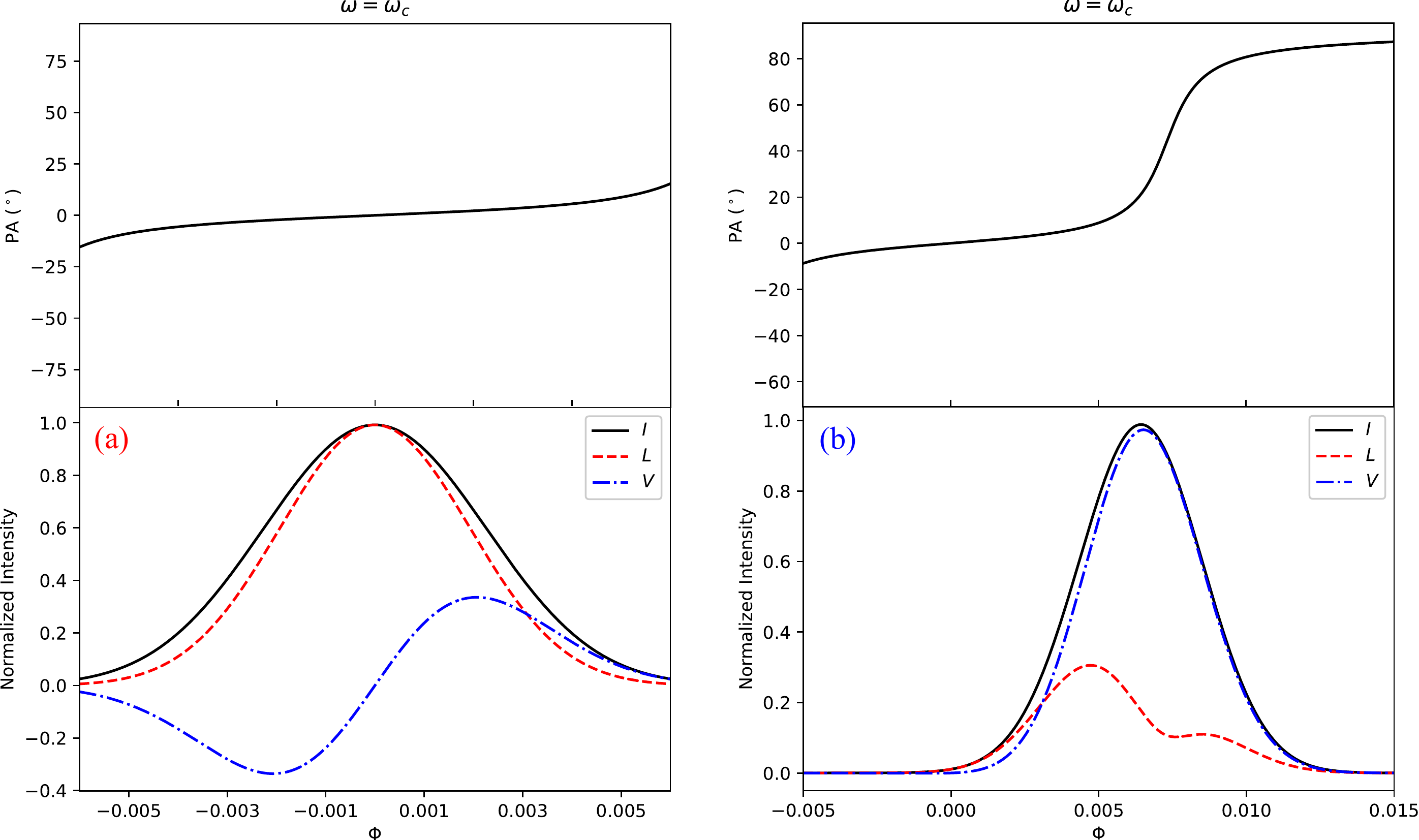}
 \caption{Simulated polarization profile and PA across the burst envelope. The total and polarized intensities are plotted in different lines: black solid line for the total, red dashed line for the linear, and blue dash-dotted line for the circular. The inclination angle and angle between the line of sight (LOS) and spin axis are adopted as $\alpha=30^\circ$ and $\zeta=45^\circ$. Parameters are quoted as $\gamma=100$, $\varphi_t=0.008$, and $\sigma_w=0.004$. Two panels have different Gaussian distributions: $\Phi_p=0$ for panel (a) and $\Phi_p=0.008$ for panel (b).
 The sign of $V$ is different from that of, e.g., \cite{Gil90} because the direction of the LOS sweeping is different. Our definition of the LOS sweeping is similar to that of \cite{Gangadhara21}.}
 \label{fig:fig2}
\end{figure}

\begin{figure}
 \centering
 \includegraphics[width = 7.5cm]{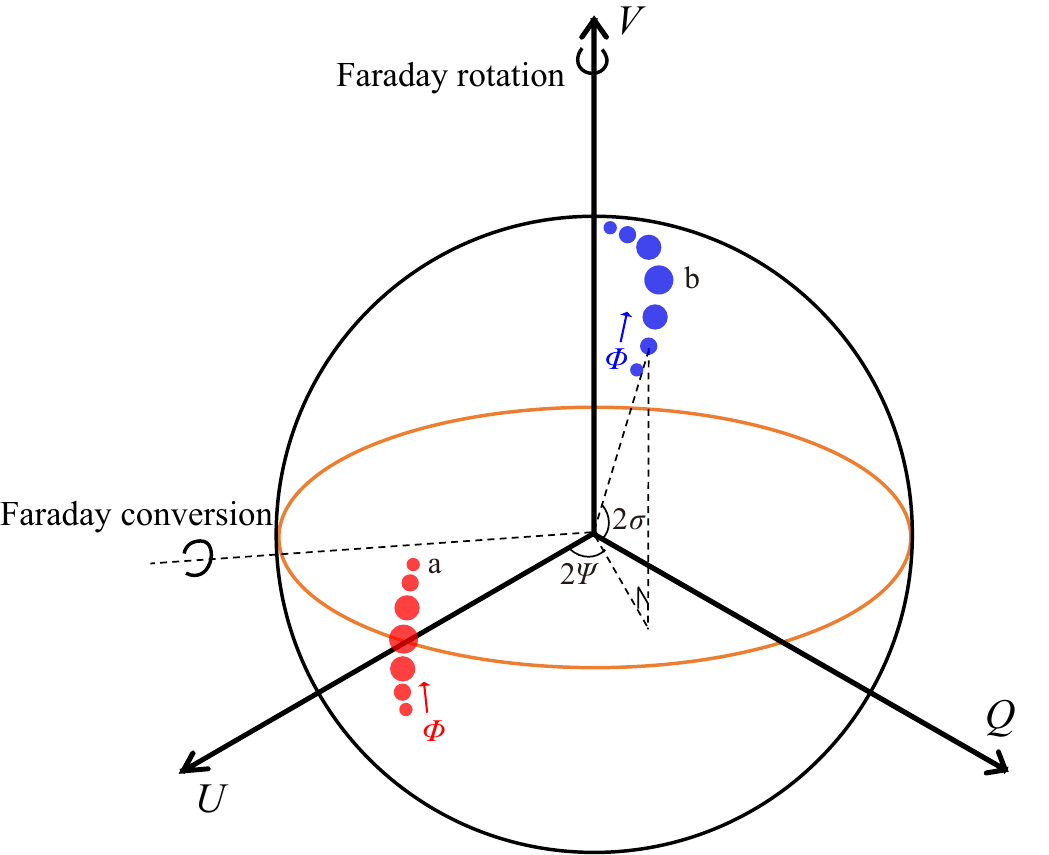}
 \caption{Schematic diagram of a bulk of Poincar{\' e} sphere. The angle $\Psi$ is the polarization position angle, and $\sigma$ is $1/2\sin^{-1}(\Pi_V)$. Colored circles denote the simulated polarization shown in Figure \ref{fig:fig2}: red circles for panel (a) and blue circles for panel (b). The area for the circles presents the value of $I$. Colored arrows show the direction of the $\Phi$ evolution, which increases in the two scenarios.}
 \label{fig:fig0}
\end{figure}

CR is generated by the perpendicular acceleration of charged particles moving along a curved trajectory.
Basically, sparking-induced charged particles can move along curved magnetic field lines in the magnetosphere, emitting CRs.
These charged particles can form a bunch if their sizes are smaller than the half wavelength of the emission \footnote{The size projected at the line of sight (LOS) should be smaller than the half wavelength, but that perpendicular to the LOS can be much larger.}.
Emitting waves are then significantly enhanced, i.e., $L=N_b N_e^2 p_e$, where $L$ is the luminosity, $N_b$ is the number of independent bunches that contribute to the observed luminosity at an epoch, $N_e$ is the number of net charge in a bunch, and $p_e$ is the emission power of an individual particle.
The critical frequency of the CR is $\nu_c=3c\gamma^3/(4\pi\rho)=0.7\gamma_2^3\rho_7^{-1}\,\rm GHz$, where $\gamma$ is the Lorentz factor of bunches, $c$ is the speed of light, and $\rho$ is the curvature radius.
To match the observed FRB frequency, the emission region should be at dozens of stellar radii with a required Lorentz factor in order of several hundred.

FRB waves will be observed if the line of sight (LOS) sweeps across the field lines where the bunches appear coincidentally.
Basically, for a single charge, $A_{\|}$and $A_{\perp}$ are the two orthogonal polarized components of the amplitude along $\boldsymbol{\epsilon}_{\|}$ and $\boldsymbol{\epsilon}_{\perp}$, where $\boldsymbol{\epsilon}_{\|}$is the unit vector pointing to the center of the instantaneous circle, and $\boldsymbol{\epsilon}_{\perp}=\boldsymbol{n} \times \boldsymbol{\epsilon}_{\|}$is defined \cite{Jackson98}.
The two amplitudes are given by \cite{Wang21}
\begin{equation}
\begin{aligned}
A_{\|} & \simeq \frac{i 2}{\sqrt{3}} \frac{\rho}{c}\left(\frac{1}{\gamma^{2}}+\varphi^{2}+\chi^{2}\right) K_{\frac{2}{3}}(\xi)+\frac{2}{\sqrt{3}} \frac{\rho}{c} \chi\left(\frac{1}{\gamma^{2}}+\varphi^{2}+\chi^{2}\right)^{1 / 2} K_{\frac{1}{3}}(\xi), \\
A_{\perp} & \simeq \frac{2}{\sqrt{3}} \frac{\rho}{c} \varphi\left(\frac{1}{\gamma^{2}}+\varphi^{2}+\chi^{2}\right)^{1 / 2} K_{\frac{1}{3}}(\xi),
\end{aligned}
\end{equation}
where $\chi$ is the angle between the considered trajectory and trajectory at $t = 0$, $\varphi$ is the angle between the LOS and trajectory plane, and $K_\nu$ is the modified Bessel function.
The shape of three-dimensional (3D) bunches can affect the spectrum and polarization.
Because the total flux for $1/\gamma < \chi$ and $1/\gamma < \varphi$ is extremely small, we only consider the case of $\chi\ll\varphi< 1/\gamma$ here.
We assume that the plasma density in the emission region is Gaussian-modulated in the azimuthal direction:
\be
f(\phi)=f_{0} \exp \left[-\left(\frac{\Phi-\Phi_{\mathrm{p}}}{\sigma_{w}}\right)^{2}\right],
\label{eq:gaussianprofile}
\ee
where $\Phi_{\rm p}$ is the peak location of the Gaussian function, i.e., at the center of the emission region, $f_0$ is the amplitude, and $\sigma_w$ is the Gaussian width.
The two orthogonal polarized amplitudes of waves are the summation of the amplitudes of individual particles \footnote{Detailed calculations are referred to Wang et al. (2021)\cite{Wang21}}:
\be
\begin{aligned}
&A_{\|,w} =\sum A_{\|}\\
&A_{\perp,w} =\sum A_{\perp}.
\end{aligned}
\label{eq:A}
\ee
By considering different 3D trajectories of bunches in the magnetosphere, the spectra of the emission can be obtained, which is generally characterized by multi-segment broken power laws \cite{Yang18}.
The model is evolved by deriving the sweeping LOS, and spectra are considered within on-beam and off-beam scenarios.
Emitted waves are slightly circularly polarized if the LOS is confined to the beam (on-beam case), and it would
become highly circularly polarized for the off-beam case. \cite{Wang21}.

The Stokes parameters for the radiation are given by
\begin{equation}
\begin{aligned}
&I=\mu\left(A_{\|,w} A_{\|,w}^{*}+A_{\perp,w} A_{\perp,w}^{*}\right) \\
&Q=\mu\left(A_{\|,w} A_{\|,w}^{*}-A_{\perp,w} A_{\perp,w}^{*}\right), \\
&U=\mu\left(A_{\|,w} A_{\perp,w}^{*}+A_{\perp,w} A_{\|,w}^{*}\right), \\
&V=-i \mu\left(A_{\|,w} A_{\perp,w}^{*}-A_{\perp,w} A_{\|,w}^{*}\right),
\end{aligned}
\end{equation}
where $\mu=\omega^2 e^2/(4\pi^2 \mathcal{R}^2 c T)$, $\mathcal{R}$ is the distance from the emitting source to the observer, and $T$ is the mean time interval of each coherent pulse.
The corresponding PA is then given by
\begin{equation}
\Psi=\frac{1}{2} \tan ^{-1}\left(\frac{U}{Q}\right)+\psi,
\end{equation}
where
\begin{equation}
\psi=\tan^{-1}\left(\frac{\sin \alpha \sin \Phi}{\cos \alpha \sin \zeta-\cos \zeta \sin \alpha \cos \Phi}\right),
\end{equation}
where $\alpha$ is the angle between the magnetic axis and rotational axis and $\zeta$ is the angle between the LOS and spin axis.
If the charges are uniformly distributed in bunches, then the summation of $A_{\perp}$ would be canceled out when the LOS sweeps at the symmetric axis of bunches.
As a result, the LP fraction is $100\%$.
If the opening angle $\varphi\ll1/\gamma$, then the CP fraction at $\omega=\omega_c$ can be estimated as
\be
\Pi_V=\frac{2\rm{Im}[A_{\|,w}]\rm{Re}[A_{\perp,w}]}{A_{\|,w}^2+A_{\perp,w}^2}\simeq\frac{2\varphi\gamma}{1+4\gamma^2\varphi^2}.
\label{eq7}
\ee
In this case, the LOS is confined to the spread angle, so the amplitude of $A_{\perp,w}$ becomes extremely smaller than that of $A_{\|,w}$.
For the off-beam case ($\varphi\gg1/\gamma$), a highly CP fraction can appear at the sides of the pulse window \cite{Wang21}. However, the flux is suppressed by a factor of $\varphi^{-2}\gamma^{-2}$.
A small $\varphi_t$ may mean a low luminosity, and bursts with small CP fractions indicate that the sparking process occurs in a small region.
Nevertheless, a low-luminous FRB could be enhanced via interstellar scintillation, and intrinsically weak bursts could be detected by this selection effect.

Luminous bursts with highly CP fractions may be produced when $\varphi_t\sim1/\gamma$ or when charges are extremely non-uniformly distributed in bunches.
Assuming that the opening angle of the bunches is $\varphi_t=0.8\gamma^{-1}$, we simulated the polarized profile and PA across the burst envelope shown in Figure \ref{fig:fig2}.
Within the pulse window, the CP fraction can reach $90\%$, matching the observations of FRB 20201124A.
Bunches with $\varphi_t\gg1/\gamma$ are strongly linearly polarized.
Consequently, the rare event rate for highly CP events may hint that most bunches have large opening angles in those cases.
Essentially, the simulated PA change across the burst arises from bunch-coherent radiation, making it different from that in the rotating vector model \cite{Radhakrishnan69}.

The simulated polarization profiles in Figure \ref{fig:fig2} are plotted in a Poincar{\'e} sphere, as shown in Figure \ref{fig:fig0}.
Both evolutionary trajectories are located at the surface of the Poincar{\'e} sphere as the emission is $100\%$ polarized.
However, evolutionary trajectories exhibit noticeable differences, showing the diversity of polarization profiles and PA across the burst envelope.
The CR model can account for diverse polarization properties, including highly linear and CPs, with either a flat or variable PA across the burst envelope.

\section{Propagation Effect}\label{sec3}

In this section, we discuss polarization properties by deriving two propagation effects, namely, cyclotron resonance absorption and Faraday conversion.
In general, plasma is thought to be cold with the magnetic field $B = (0,\,B_y,\,B_z)$.
The wave vector $k$ is parallel to the $z$-axis.

\subsection{Cyclotron Resonance Absorption}\label{sec3.1}

\begin{figure}
 \centering
 \includegraphics[width = 7.5 cm]{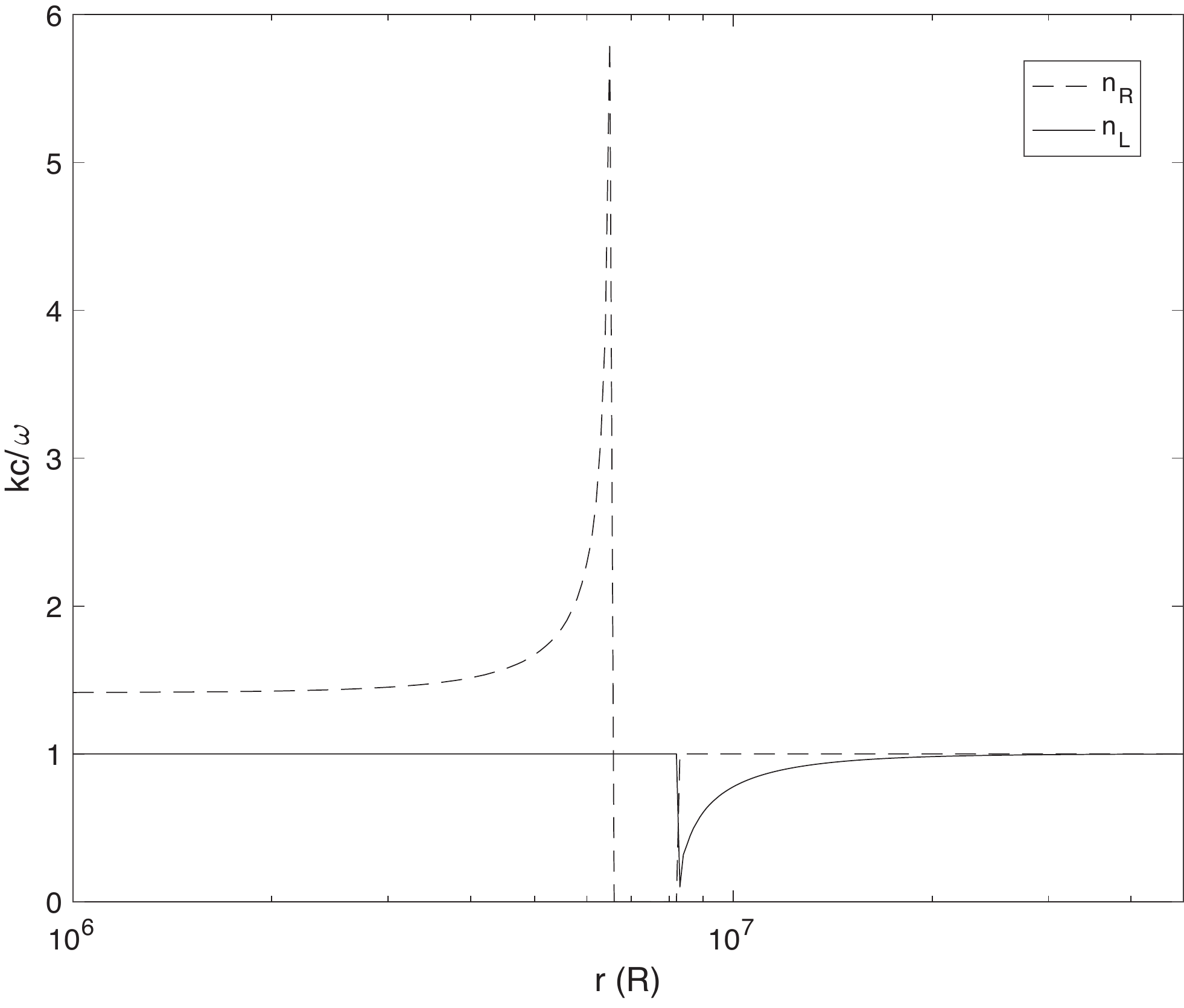}
 \caption{Dispersion relationship as a function of height. Parameters are adopted as $B_s=10^{12}$ G, $P=0.01$ s, and $\omega=2\pi\times10^9$ Hz. Photons with different modes are presented as L-mode (black solid line) and R-mode (black dashed line). The height is in the unit of the stellar radius.}
 \label{fig:fig3}
\end{figure}

\begin{figure}
 \centering
 \includegraphics[width = 7.5 cm]{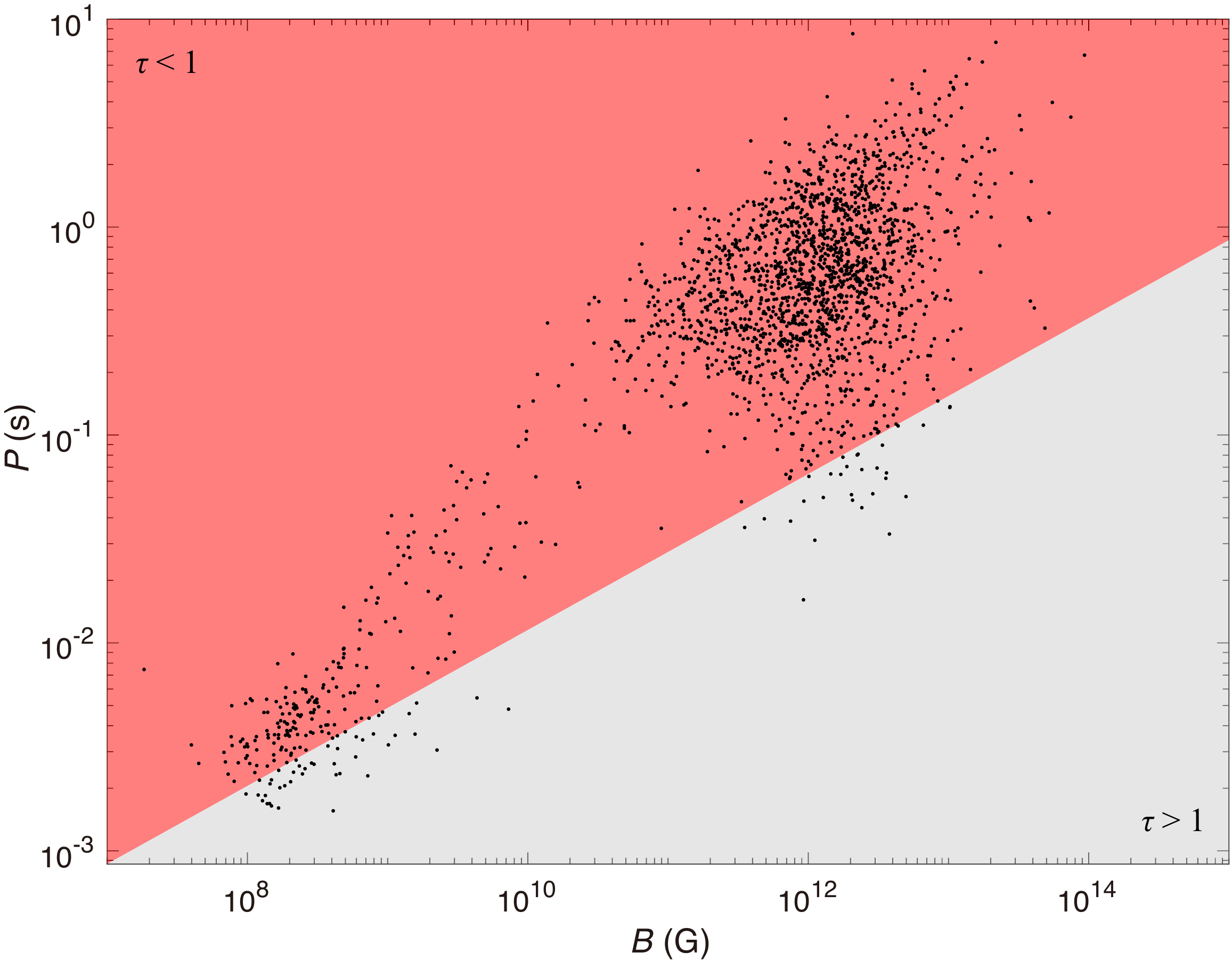}
 \caption{Regions of the optical depth for R-mode photons: gray region (optically thick for $\tau>1$); red region (optically thin for $\tau<1$). Black dots present pulsars, where $B$ is the surface magnetic field. Data are quoted from Manchester et al. (2005) \cite{Manchester05}. }
 \label{fig:fig4}
\end{figure}

In this subsection, we discuss the propagation of waves in the linear regime.
We consider that the propagation effect occurs at the closed field lines.
The dispersion relationship is given as \cite{Meszaros92}
\be
k^2c^2=\omega^2\left[1-\frac{\omega_{\mathrm{p}}^{2}}{\omega^{2}} \frac{1}{1-\frac{1}{2}
\frac{\omega_{y}^{2}}{\omega^{2}-\omega_{\mathrm{p}}^{2}} \mp \sqrt{\frac{\omega_{z}^{2}}{\omega^{2}}+\left(\frac{1}{2} \frac{\omega_{y}^{2}}{\omega^{2}-\omega_{\mathrm{p}}^{2}}\right)^{2}}}\right],
\label{eq:nk}
\ee
where $\omega_{y}=(B_{y}/B)\omega_{B}$, $\omega_{z}=(B_{z}/B) \omega_{B}$, $\omega_B$ is the
cyclotron frequency, and $\omega_p=(4\pi n_e e^2/m_e)^{1/2}$ is the plasma frequency.
Generally, the number density is $n_e=\kappa n_{\rm GJ}$, where $n_{\rm GJ}= |\boldsymbol{\Omega}\cdot\boldsymbol{B}|/(2\pi ce)$ is the Goldreich-Julian density \cite{GJ69}.\
In the closed-field-line region, there is no outward charge flow, and the charge density can be described by the static charge density (e.g., $\kappa=1$).
The number density is likely complicated for magnetars, but the exact density is difficult to infer from observations.
For simplicity, we use $n_{\rm GJ}$ as a proxy of the number density.
Although some models (e.g., \cite{Lyutikov08}) proposed that the number density of magnetars exceeds that of the pulsar by a factor of $10^4$, it does not affect the dispersion relationship.

In Equation (\ref{eq:nk}), the radiation propagating along the $z$-axis has two modes: L-mode and R-mode\footnote{Either of the two modes is an elliptical polarization mode.
The alphabet R(L) refers to the wave being a right (left) circularly
polarized wave for a positive $Bz$. The eigenmodes are marked as ``+'' for the L-mode and ``$−$'' for the R-mode.}.
To allow the wave to propagate in the plasma, $k^2c^2/\omega^2>0$ should be satisfied.
According to Equation (\ref{eq:nk}), L-mode waves can freely propagate, while R-mode waves would be absorbed when
$$
\omega^{2}\left(1-\frac{\omega_{p}^{2}}{\omega^{2}} \frac{\omega^{2}-\omega_z^{2}}{\omega^{2}}\right)=\omega_{B}^{2}.
$$
The waves can propagate again when
\begin{equation}
(1-r)-\sqrt{\frac{\omega_z^{2}}{\omega^{2}}+r^{2}}=\frac{\omega_{p}^{2}}{\omega^{2}},
\end{equation}
where
\begin{equation}
r=\frac{1}{2} \frac{\omega_y^{2}}{\omega^{2}-\omega_{p}^{2}}.
\end{equation}
For a typical neutron star, the condition that $\omega_p\ll\omega\ll\omega_B$ is always satisfied within the magnetosphere.
The absorption condition can be approximated as $\omega\simeq\omega_B$, and the waves can propagate again when $\omega\simeq\sqrt{\omega_B^2+2\omega_p^2}$.

At the absorption region, the electric vector for L-mode waves is given as
\be
E_{y}=-i\frac{\omega}{\omega_z}E_x.
\ee
Therefore, the CP fraction is
\be
\Pi_V=\frac{2B_zB}{B^2+B_z^2}.
\ee
The outflow radiation is elliptically polarized and becomes highly circularly polarized when $B_z\simeq B$.
If there are L- and R-mode photons, then the emission would also be elliptically polarized.

We consider a dipolar magnetic configuration ($B=B_sR^3/r^3$), where $B_s$ is the magnetic field strength at the stellar surface and $R$ is the stellar radius.
The refractive indices for R-mode waves with different altitudes are shown in Figure \ref{fig:fig3}, with the assumption of $B_s=10^{12}$ G, $P=0.01$ s, and $\omega=2\pi\times10^9$ Hz.
When R-mode waves propagate along the absorption region, the optical depth can be calculated as \cite{Wang10,Lu21}
\be
\tau=\int i2kdz\simeq-4 \pi^{2} n_{e} e \frac{1+\frac{3 \omega_z^{2}}{\omega_y^{2}}+\frac{\omega_z^{4}}{\omega_y^{4}}}{\left(1+\frac{\omega_z^{2}}{\omega_y^{2}}\right)^{\frac{3}{2}}} \frac{1}{\frac{\mathrm{d} B_{y}}{\mathrm{~d} r}}\simeq1.7\times10^{-1}P_0^{-1}B_{s,12}^{1/3}\omega_9^{-1/3}.
\ee
The radiation would be significantly absorbed if the optical depth is larger than 1.
We plot the regions of the optical depth with different surface magnetic fields and periods and some pulsars and magnetars in Figure \ref{fig:fig4}.
The surface magnetic field is the characteristic magnetic field, which is estimated under the assumption that a pulsar spin-down is dominated by dipolar magnetic
braking.
R-mode waves are likely to be optically thick for a rapidly rotating neutron star with a strong magnetic field.
In this case, emissions with highly LP fractions would be emitted from high-altitude regions (at least higher than the absorption region).

\subsection{Faraday Conversion}\label{sec3.2}

\begin{figure}
 \centering
 \includegraphics[width = 7.5 cm]{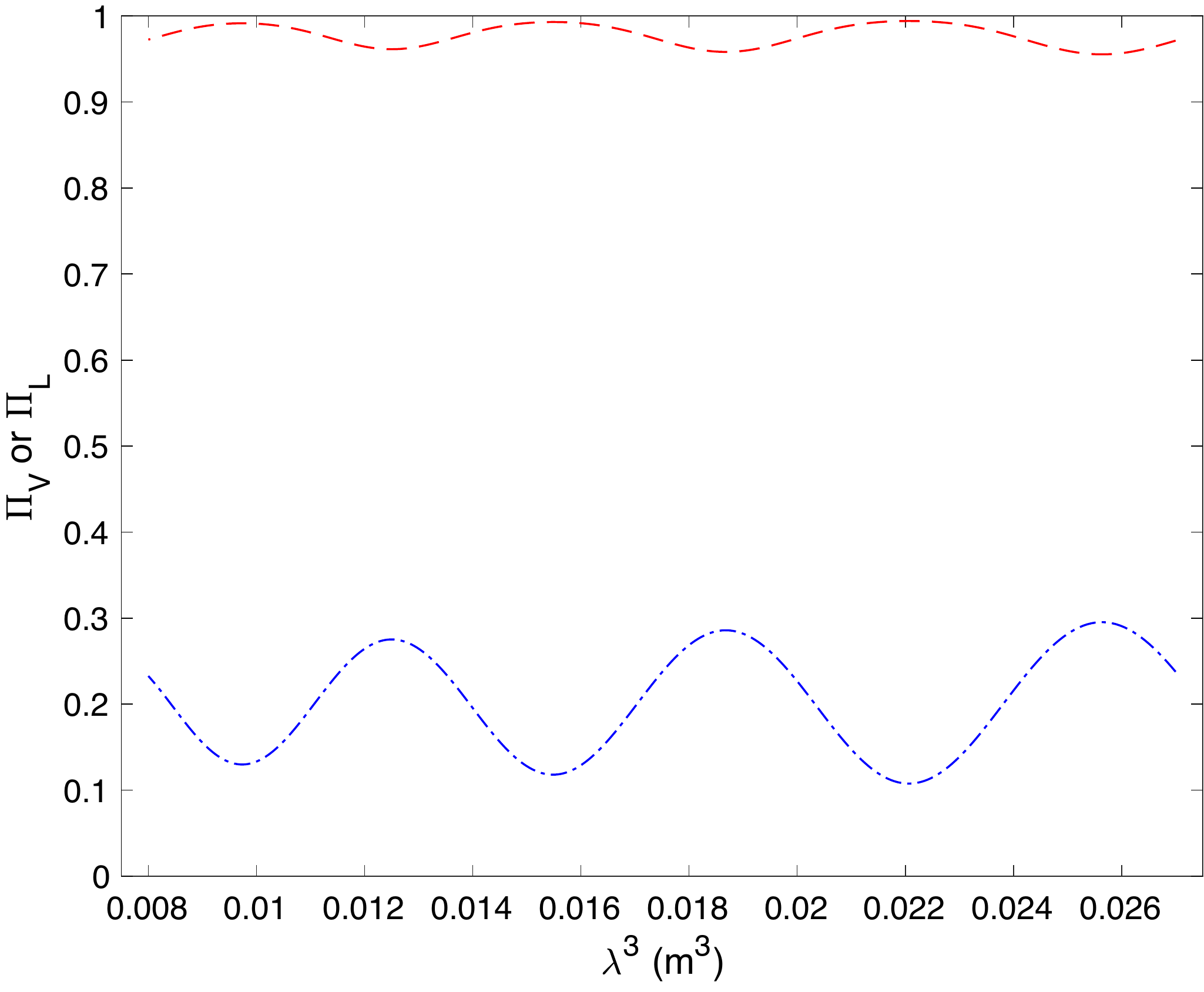}
 \caption{Simulated linear and CP fractions. The emission is $100\%$ polarized. Linear (red dashed line) and circular (blue dotted-dashed line) polarization fractions show $\lambda^3$ oscillation from the Faraday rotation. The income wave has a PA of $0$ and $\Pi_{V,0}=0.2$. Parameters are adopted as $B=100\,\rm G$, $n_eL=10^{13}\,\rm cm^{-2}$.}
 \label{fig:fig5}
\end{figure}

Another propagation effect responsible for producing the
measured CP of FRBs is suggested to be the Faraday conversion between the linear and CPs \cite{Gruzinov19,Vedantham19}.
Assuming that the plasma is cold and uniformly magnetized $\boldsymbol{B}=(0,B_y.B_z)$, the Stokes parameters can change in the Faraday screen.
As shown in Figure \ref{fig:fig0}, the Faraday conversion would happen when the Stokes parameters rotate around the $V$ axis\footnote{Faraday rotation would happen when the Stokes parameters rotate around an axis in the equatorial plane.}.
If the ratio of the Faraday conversion rate to that of the Faraday rotation is $h/f=\omega_B/\omega>1$ \cite{Sazonov69}, then the Faraday conversion will become significant.
For a radio wave with $\omega=2\pi\times10^9$ Hz, we have $B\simeq3.6\times10^2$ G.
It requires an extremely magnetized Faraday screen around the FRB source.

For FRB 20201124A, the observed DM is $\rm DM\approx410\,pc\,cm^{-3}$, which lets the Faraday conversion screen of $n_e L\lesssim\rm10^{21}\,pc\,cm^{-3}$, where $L$ is the spatial scale of the screen.
With the assumption of $B=100$ G, $n_e L\simeq3\times10^{13}\,\rm pc\,cm^{-3}$ should be satisfied to sustain a measured $\rm|RM|\simeq10^3\,m^{-2}$.
We simulated the evolution for the linear and CP fractions, shown in Figure \ref{fig:fig5}.
The linear and CP fractions oscillate with $\lambda^3$, but their average values only depend on the income waves.
However, the observations of FRB 20201124A show a $\lambda^2$-oscillation for Stokes parameters \cite{Xu21b}, which is inconsistent with the Faraday conversion in a cold plasma.
A high magnetic field strength can enhance the amplitude of the oscillation, but it also makes the oscillation occur frequently.
As a result, outcome emissions with a highly CP need the income waves to have large CP fractions.

\section{Discussion}
\label{sec4}

\subsection{Coherent Inverse Compton Scattering by Bunches}\label{sec4.1}

Motivated by an explanation of the observed core and conal emission beams~\cite{Rankin93}, Qiao \& Lin~\cite{Qiao98} proposed an inverse Compton scattering (ICS) model of the radio pulsar radiation.
A neutron star may undergo a sudden trigger, and there are some low-frequency electromagnetic waves emitted by the sparking from the inner polar cap region (Fig.~\ref{fig:fig1}).
The waves enter charged bunches and are then scattered by the relativistic particles.
The frequency of the scattered waves is
$
\omega_{\text {out }}=1.5 \omega_{\text {in }} \gamma^{2}\left(1-\beta \cos \theta_{\text {in }}\right)
$,
where $\omega_{\text{in}}$ is the angular frequency of an incident photon, $\beta$ is the dimensionless velocity of a particle, and $\theta_{\text{in}}$ is the angle between the radial direction and tangent direction of magnetic field lines (inclination angle).
To create GHz-FRB emissions, $\omega_{\text {in}}\sim10^5\,\rm{rad\,s^{-1}}$ is required.
Within the physical picture of the ICS mechanism, particles are accelerated by the oscillating electromagnetic field of low-frequency waves.

In the regime of the ICS model, Xu \& Qiao~\cite{Xu98} considered the coherent radiation of charged bunches and argued that ``the CP intensity of radio pulsar's emission might be the result of coherent superposition of an emission bunch.''
An observer can see the scattered emission by many bunched particles around the LOS.
The outcome wave is linearly polarized after the ICS by a single particle, but the electric complex vectors of the scattered waves in the same direction are added for all the particles in a bunch.
The total Stokes parameters can then be obtained from the total complex vector for this bunch.
Under the assumption of a constant Lorentz factor at a certain altitude, the simulated polarized profile and PA across the burst envelope are shown in Liu et al. (1999)~\cite{Liu99} and Xu et al. (2000)~\cite{Xu00} for single pulses (a strong likeness to the FRBs) and integrated profile.
Similar to the CR scenario, the sign of the CP fraction would change, and pulses have a $\sim 100\%$ LP fraction when the LOS sweeps at the bunch's center.
This finding is not surprising because both ways of coherent superposition are relevant to a bunch of charged particles but only different in the radiation mechanism (either CR or ICS). Thus, a circularly polarized emission can naturally appear by adding up so many linearly polarized waves with different phases and electric vectors produced by every particle.
Recently, Zhang~\cite{Zhang21} proposed a family of models invoking the coherent ICS of bunched particles that may operate within or just outside of the magnetosphere of a flaring magnetar.

Bunched particles are ultra-relativistic for CR and ICS scenarios.
For such particles, the radiation is beamed in a narrow cone, which has an opening angle of $\sim 1/\gamma$ when $\omega=\omega_c$ in the direction of the particle's velocity.
Within the scenario of ICS, circularly polarized emissions from bunches are produced because linearly polarized waves from particles are added with different phases and vectors.
However, CR can be highly circularly polarized at the wing of the radiation beam (e.g., see~\cite{Tong22}, which scenario is similar with CR from a single charge).
If the bunches' opening angle $\varphi_t\gg1/\gamma$, emission would be highly linearly polarized since L- and R-mode waves cancel each other out \cite{Wang21}.

\subsection{Production of a bulk of energetic bunches}

We have demonstrated that linearly and circularly polarized emissions can be created through coherent radiation of bunches. However, how can a bulk of large bunches with energetic particles be produced near the surface of a compact star in its polar cap region?
This is essentially a question relevant to the nature of pulsar surface and even of cold dense matter at a supra-nuclear
density~\cite{Xu2020}.
Historically, the ultra-relativistic particle cascade, i.e., the sparking in the gap near the pulsar surface, was proposed in the 1970s, but the long-standing binding energy problem in the Ruderman–Sutherland~\cite{RS75} scenario could be solved if there is an extremely strong magnetic field on the pulsar's surface with a moderate temperature \cite{Medin07},
or if strongly bound free quarks are supposed to exist on the surface \cite{Xu99}.
Furthermore, the strong coupling between quarks may render a few quarks grouped together to form the so-called strange matter in a solid state because of the color interaction in between at the {\em low-energy} regime~\cite{Xu03}.
The high-tension point discharges on the rough solid surface may help trigger several bunches of electron-positron pairs, during a quake-induced and oscillation-driven magnetospheric activity for instance~\cite{Lin15}, though the development process of positively (or negatively, depending on the angle between rotational and magnetic axes) charged bunches awaits further study in the future.
Several bunches depicted in Fig.~\ref{fig:fig1} could be generated during the oscillations of a sudden quake. However, only one of the coherent emissions in some magnetic flux tubes, i.e., the bulk of a few bunches in the two thick lines of Fig.~\ref{fig:fig1}, would be detected by an observer.
In fact, a single pulse observation with China's FAST may hint at a rough surface of PSR B2016+28 \cite{Lu19}.
In this sense, a comprehensive study of repeating FRBs is surely meaningful for the radiative mechanism of the coherent emission of radio pulsars and for the physics of a superdense matter.

{\em What is the magnetospheric difference between a regular pulsar and repeating FRB} if pulsar-like compact stars are responsible for both kinds of the extremely coherent radio emission?
The essential difference may arise from the fact that regular pulsars are basically rotation powered, while repeating FRBs could probably be powered by a stellar activity via either a starquake or magnetic reconnection near the surface \cite{Wang18,Wadiasingh19}.
An FRB object is usually below the deathline at ordinary times, exhibiting a vacuum-like clean magnetosphere, but electronic plasma can occasionally erupt from the star during an active period.
However, a radio pulsar is always surrounded by a relatively dense plasma because of the high potential drop in the open field line region.
This may result in a higher coherence and thus a bright emission of FRBs.

\subsection{Repeating and apparently non-repeating FRBs}

Repeaters seem to exhibit noticeably three properties different from apparently non-repeating FRBs.
(1). As is well discussed, repeating sources seem to have less luminous bursts than most apparently non-repeating ones \cite{Luo20b}.
(2). The time--frequency drifting pattern has been detected in at least some of the repeater's subpulses but not for apparently non-repeating FRBs.
An apparently non-repeating source, i.e., FRB 181123, shows a frequency pulse-to-pulse downward evolution, but the interval for the pulses is much larger than burst duration \cite{Zhu20}.
%
(3). Polarization properties between the two classifications are also different, and several apparently non-repeating FRBs have polarization measurements.
The LP percentage for the apparently non-repeating FRBs ranges from $8.5\%$ to $80\%$ and from $3\%$ to $23\%$ for the CP \cite{Masui15,Petroff15,Caleb18}.
The total polarization degree for such apparently non-repeating FRBs is much less than $100\%$, which is different from some repeaters \cite{Michilli18,CHIME20,Xu21a}.

One possibility is that apparently non-repeating FRBs with a low degree of polarization may hint radiation to be less coherent than that of the repeater's bursts if all FRBs share a similar radiation mechanism, e.g., coherent emission by bunches.
The high luminosity of non-repeating ones requires a much more number of emitting charges, and if so, then low-frequency emissions might be optically thick in this case.
The other possibility is that apparently non-repeating FRBs originate from quite different energy-providing processes, preferring catastrophic events~\cite{Jiang20}, but it is not sure if the coherent radiation mechanisms of both types of FRBs are similar.

\section{Summary}
\label{sec5}

We investigate the possible processes of producing highly CP by deriving the intrinsic mechanism and propagative effect.
The intrinsic mechanisms by invoking charged bunches are mainly discussed in the coherent CR scenario, although it is similar in a coherent ICS model.
When the opening angle $\varphi_t\sim\gamma^{-1}$ or particles are extremely non-uniformly distributed in bunches, coherent CRs from the bunches can be highly circularly polarized.
The mechanisms by invoking charged bunches can also interpret the narrowband spectrum and downward-drifting structure of repeating FRBs \cite{Yang18,Wang19,Wang21,Zhang21}.

Propagation effects, including cyclotron resonance absorption and Faraday conversion, are investigated to interpret the highly CP.
If the emission region is at a low altitude of the magnetosphere, R-mode photons would be absorbed, while L-mode photons can propagate, which exhibits a highly CP.
The absorption is significant, especially for fast-rotating magnetars.
If so, then bursts with strong LP fractions should be emitted at high regions, which may require an extra trigger mechanism.
A highly magnetized cold plasma region can convert LP fractions to CP fractions.
Linear and CP fractions oscillate with $\lambda^3$, and the average value of the oscillations only depends on the income waves.
This mechanism may generate CP, but the wave with a highly CP requires a highly circularly polarized income wave.

\end{CJK*}

\vspace*{2mm} \Acknowledgements{\bahao  }
We are grateful to Prof. Bing Zhang for helpful comments and discussions.
This work is supported by the National Key R\&D program of China No. 2017YFA0402602, National SKA Program of China No. 2020SKA0120100 and the strategic Priority Research Program of CAS (XDB23010200).
W.-Y.W. is supported by a Boya Fellowship and the fellowship of China Postdoctoral Science Foundation No. 2021M700247.



\bibliographystyle{unsrt}





\end{document}